# Resonant Transmission Line Method for Econophysics models

Theophanes E. Raptis

Computational Applications Group, Division of Applied Technologies, National Center for Science and Research "Demokritos"

Email: rtheo@dat.demokritos.gr

**Abstract**: In a recent paper [1304.6846], Racorean introduced a formal similarity of the Black-Sholes stock pricing model with a Schrödinger equation. We use a previously introduced method of a resonant transmission line for arbitrary $2^{nd}$ order Sturm-Liouville problems to attack the same problem from a different perspective revealing some deep structures in the naturally associated eigenvalue problem.

1. Introduction

Recently, Racorean [1] provided an example of reduction for a particular problem in actuarial mathematics into the Sturm-Liouville paradigm via a transformation of the well known, Black-Sholes equation [2] for stock pricing. Despite some difficulties in the physical interpretation of the resulting model, we take this problem as an abstract example that can be analyzed in its full generality via a different methodology and show that it can be directly associated with a double eigenvalue problem on a Lennard-Jones potential. To this aim we use a recently revived [3] approach for the Sturm-Liouville eigenvalue problem first presented by Papageorgiou at the 1980s [4], [5] based on a recursive impedance map. Care should be taken with respect to any physical interpretations of the resulting eigenvalues and eigenfunctions in the real stock market activity which is of course a man made emergent phenomenon and not a strictly low level physical system.

The general method for locating eigenvalues this way appears associated with a particular theorem which has yet to be proved analytically for this class of maps. A wide literature review provides evidence for associating this method with a previously introduced class of analytical functions known as $m(\lambda)$ by Weyl and Titchmarsh at the 1930s [6], [7]. An interesting connection after separating real and imaginary parts can also be made with so called, Birational Projective Geometries dealing with rational transformations of certain varieties [8], [9]. Their significance stem from the existence of distinct group structures known as the Cremona and De Jonquieres groups [10], [11] containing highly nontrivial structures. These have recently been proven to be not simple as an abstract group by Cantat and Lamy [12]. Notably, by a theorem due to Noether we know that the general case of Cremona maps admits a factorization into a product of simpler De Jonquieres maps for which there also exists a resolution into quadratics [13]. We should also mention here some recent attempts by Saniga [14] to associate such structures with the deeper substrate of macro-physical space-time as well as with certain phenomena of conscious experience of temporal and spatial dimensions [15], [16].

While the above call for a deeper study of the particulars of the class of maps and dynamical systems defined by the transmission line (TL) model, in the next section we restrict attention to the case of the original Black-Sholes model and provide a set of transformations representing a natural mapping to the standard Sturm-Liouville form with two parameters. In section 3, we explore the resonant TL model for locating eigenvalues and eigenfunctions in the positive part of the two parameters plane.

## 2. Transformation of Black-Sholes to an eigenvalue problem

Based on the material presented in [2], we proceed with the appropriate identification of terms in the Black-Sholes model. Using the general form

$$\partial_t V + KS^2 \partial_S^2 V + rS\partial_S V - rV = 0, \quad K = \sigma^2/2 \qquad (1)$$

In the above $V(S, t)$ is the option price as a function of both time and the stock price volatility $\sigma$ with a risk-free interest rate, which is assumed to factorize a multifractal Brownian process $Ds/s = \mu Dt + \sigma dW$ of volatility $\sigma$ as $\exp(-rt)V(S,t)$. Using standard protocol for variable separation we get the time dependence as $v(t) = \exp(-rt/\sigma)$ and the stock price dependence as

$$2S^2 \partial_S^2 V + r\sigma S\partial_S V + r(1-\sigma)V = 0 \qquad (2)$$

We can turn (2) into a Sturm-Liouville form in order to extract the associated TL system after first rationalizing the expression with the transformation $(uv)'' = u''v + 2u'v' + uv''$. Inserting a new variable $W(S) = f(S)V(S)$ allows writing (2) in the alternative form

$$W'' + P(S)W' + Q(S)W = 0 \tag{3a}$$

Using $V' = (W'/f) - (\log f)'(W/f)$ and $2S^2V'' = (2S^2/f)W''$ in (2) we find

$$P(S) = r\sigma - \frac{(\log f(S))'}{S}$$
$$Q(S) = \frac{r(1-\sigma)}{f(S)} - r\sigma f'(S) + \frac{|f'(s)|^2}{Sf(S)} \tag{3b-c}$$

We then turn the above to a double eigenvalue problem with the choice $f(S) = kS, k > 0$ such that (3a-c) becomes

$$W'' + \left(\frac{\lambda}{k} - \frac{k}{S^2}\right)W' + \left(\frac{k}{S^2} - \frac{\mu}{k^2 S} - \lambda\right)W = 0 \tag{4a}$$

$$\lambda = kr\sigma, \mu + \lambda = kr \tag{4b}$$

We can now complete the identification of (3a-b) with a resonant transmission line using the exact same method introduced in [2], section 2. Specifically, we first turn (4) into the equivalent Sturm-Liouvile form as

$$k\frac{d}{ds}\left[e^{\lambda ks + \frac{1}{ks}} k\frac{dW}{ds}\right] + e^{\lambda ks + \frac{1}{ks}}\left(\frac{k^3}{s^2} - \frac{\mu}{s} - \lambda\right)W = 0 \tag{5}$$

In (5) we used the rescaled variable $s = S/k^2$.

### 3. The resonant transmission line method

The corresponding 1$^{st}$ order ODE system for the representation of a 4-port T-quadrupole circuit in the relevant TL model will be given as

$$\begin{cases} dV(x)/dx = -\mathbf{i}\gamma^2(x,\mu,\lambda)e^{\beta(x,\lambda)}I(x) \\ dI(x)/dx = -\mathbf{i}e^{-\beta(x,\lambda)}V(x) \end{cases} \qquad (6\text{a-b})$$

In the above system, we identify the corresponding elements for the effective potential and the propagation factor as

$$x = s, I(x) = W(s), \beta(x,\lambda) = \lambda kx + 1/kx$$

$$\gamma(x,\mu,\lambda) = \frac{\mathbf{i}}{k}\sqrt{\frac{k^3}{x^2} - \frac{\mu}{x} - \lambda} \qquad (7)$$

In what follows we may safely assume the volatility $\sigma$ sufficiently larger than unity so that both $\mu$ and $\lambda$ will also be positive. We notice a similarity of the above with the well known n$^{th}$ order interatomic Lennard-Jones potential $\varepsilon[(r_0/r)^{2n} - 2(r_0/r)^n]$ for $n = 1$, bonding energy $\varepsilon = 1$ and interatomic distance taken as $r_0 = \mu/2 = k^{3/2}$. The roots of the quadratic are given as $-(\mu/2\lambda)(1 \pm \sqrt{1 + k\mu/2})$ so that the eigenvalues ratio adjusts the width of the region with possible bound states. In what follows we may put $k = 1$ as it is an arbitrary parameter. Deep potential wells are in the region $\mu \gg \lambda$.

Applying the same technique as in [2] requires the values of all parameters to be fine tuned so as the TL to remain resonant across the x-axis which is equivalent to the condition that the input and output reactances remain matched. For an effective potential with bound states there will be certain inversion points so that this will only be possible for certain values of the parameters thus defining the problem eigenvalues via a root finding algorithm. This is given via a recursive procedure as the condition

$$\text{Im}(Z(0)) + \text{Im}(G_{CFE}^n(Z_B(\delta x), Z_P(\delta x), Z(L))) = 0, \quad n = L/\delta x \qquad (8)$$

In (8), $G_{CFE}$ is a continued fraction expansion operator acting on remote input impedances at a distance $L$ that has to be matched with the impedance at 0 starting point. In our case, there is always a pole at zero so that there is an equivalent infinite load with a +1 reflection coefficient

(open circuit standing wave). The characteristic impedances for (8) are given via the approximation $|\gamma(x)|\delta x \ll 1$ as

$$Z_B = Z\tanh(\gamma e^\beta dx/2) \approx -i\gamma^2(x,\mu,\lambda)e^{\beta(n\delta x,\lambda)}\delta x/2$$

$$Z_P = \frac{Z}{\sinh(\gamma e^\beta dx)} \approx ie^{\beta(n\delta x,\lambda)}/\delta x \qquad (9)$$

$$Z = \gamma(x,\mu,\lambda)e^{\beta(n\delta x,\lambda)}$$

Simplifying the resulting expression gives a dynamical system of the form

$$Z_{n+1} = \frac{\gamma_R \delta x + i(\gamma_I + 2Z_n e^{-\beta})}{2\left[(Z_n \delta x + \gamma_I e^\beta \delta x^2) + ie^{-\beta}\left(1 - \gamma_R e^{2\beta}\delta x^2\right)\right]} - i\gamma e^\beta \delta x/2 \qquad (10)$$

In (10) we used $\gamma_R = \mathrm{Re}(\gamma), \gamma_I = \mathrm{Im}(\gamma)$ to emphasize the fact that this quantity alternates when passing through inversion points of the potential for specific eigenvalues with one of the two terms vanishing. The same procedure can be repeated for the whole of the $(\mu, \lambda)$ upper positive plane and specific eigenvalue pairs can be selected for any particular choice of $k$, $r$ and $\sigma$ parameters via (4b). The roots of (8) can also be traced as the absolute minima at 0 of the resulting surface $\|\mathrm{Im}(Z(0) + Z_n(\mu,\lambda))\|^2$. In fact we may write (10) in the form of an abstract 2-dimensional nonlinear dynamical system comprising a birational map of the plane by separating real and imaginary parts as

$$X_{n+1} = \frac{a(x_n)X_n + b(x_n)Y_n + K_1(x_n)}{G(X_n, Y_n, x_n)}$$

$$Y_{n+1} = \frac{c(x_n)X_n + d(x_n)Y_n + K_2(x_n)|Z_n|^2 + K_3(x_n)}{G(X_n, Y_n, x_n)} \qquad (11)$$

$$G(X_n, Y_n, x_n) = |Z_n|^2 \delta x^2 + g_X(x_n)X_n + g_Y(x_n)Y_n + K_4(x_n)$$

$$|Z_n|^2 = X_n^2 + Y_n^2$$

Exact expressions of the coefficients in (11) can be readily found with any computer algebra system but they are too complicated to reproduce in here. The essence of the algorithm is that there will be a specific subset of isolated points in the complex plane corresponding to the original problem eigenvalues for which the recursion over (11) returns to the imaginary axis at $X_n = 0, Y_n = Y(0)$ when started from a remote point at $x = L$ with $X_0 = 0, Y_0 = Y(L)$ thus guaranteeing the resonant, "lossless" character of the TL model which is the necessary condition allowing for sustainable waves. Eigenfunctions can be readily obtained as soon as the eigenvalues are known by the direct use of a transfer matrix as predicted from the general solution of the telegrapher's equation.

Numerically, the particular potential shows great instabilities and special effort is required for variable step methods in order to satisfy the necessary condition $|\gamma(x)| \delta x \ll 1$. Details of numerical experimentation will be reported elsewhere.

## 4. Discussion and Conclusions

We examined the case of Black-Sholes equation as an abstract example for the application of a previously introduced transmission line method for eigenvalue problems. The method appears promising while the exact expressions of the recursive scheme used show an intricate underlying geometry that deserves further examination. Regarding econophysical applications, interpretation of results should always be taken in a different context than that of abstract physical systems. Specifically, we refer the reader to a recent study by Derman [17] where the difference between strictly physical models and social ones is sufficiently stressed based on the fact that the former are immutable while the latter present an additional external instability in so far as they are amenable to psychological and other variations not included in the models themselves. Possibility of improving models by including such external factors is perhaps in the realm of present technological capacities using some form of multi-physics simulations including general automata and parallel computations. Recent examples include the so called, "Sentient World Simulation" by Chaturvedi [18] developed for terrorism studies.